\documentclass[%
preprint,
 amsmath,amssymb,
 aps,
]{revtex4-1}

\usepackage{graphicx}
\usepackage{dcolumn}
\usepackage{bm}
\usepackage{xcolor}
\usepackage{hyperref}
\usepackage{qcircuit}
\usepackage{braket}
\usepackage{multirow}

\renewcommand{\figureautorefname}{Figure~\negthinspace}

\renewcommand{\tableautorefname}{Table~\negthinspace}

\begin{document}

\preprint{BNL}

\title{Federated Quantum Machine Learning}

\author{Samuel Yen-Chi Chen}
\email{ychen@bnl.gov}
\affiliation{%
 Computational Science Initiative, Brookhaven National Laboratory, Upton, NY 11973, USA
}%
\author{Shinjae Yoo}%
 \email{sjyoo@bnl.gov}
\affiliation{%
 Computational Science Initiative, Brookhaven National Laboratory, Upton, NY 11973, USA
}%




\date{\today}

\begin{abstract}
Distributed training across several quantum computers could significantly improve the training time and if we could share the learned model, not the data, it could potentially improve the data privacy as the training would happen where the data is located. 
However, to the best of our knowledge, no work has been done in quantum machine learning (QML) in federation setting yet.
In this work, we present the federated training on hybrid quantum-classical machine learning models although our framework could be generalized to pure quantum machine learning model. Specifically, we consider the quantum neural network (QNN) coupled with classical pre-trained convolutional model. 
Our distributed federated learning scheme demonstrated almost the same level of trained model accuracies and yet significantly faster distributed training. It demonstrates a promising future research direction for scaling and privacy aspects.
\end{abstract}

\maketitle


\section{\label{sec:Indroduction}Introduction}
%
Recently advances in machine learning (ML), in particular deep learning (DL), has found significant success in a wide variety of challenging tasks such as computer vision \cite{simonyan2014very, Szegedy2014GoingConvolutions, Voulodimos2018DeepReview}, natural language processing \cite{Sutskever2014SequenceNetworks}, and even playing the game of Go with superhuman performance \cite{Silver2016MasteringSearch}. 

%
In the meantime, quantum computers are introduced to the general public by several technology companies such as IBM \cite{cross2018ibm}, Google \cite{arute2019quantum}, IonQ \cite{grzesiak2020efficient}
and D-Wave \cite{lanting2014entanglement}. Theoretically, quantum computing can provide exponential speedup to certain classes of hard problems that are intractable on classical computers \cite{harrow2017quantum, nielsen2002quantum}. The most famous example is the factorization of large numbers via Shor algorithm \cite{shor1999polynomial} which can provide exponential speedup. While the search in unstructured database via Grover algorithm \cite{grover1997quantum} can provide quadratic speedup. However, currently available quantum computers are not equipped with quantum error correction \cite{gottesman1997stabilizer,gottesman1998theory} and would suffer from the device noise. Quantum computation tasks or quantum circuits with a large number of qubits and/or a long circuit depth cannot be faithfully implemented on these so-called noisy intermediate-scale quantum (NISQ) devices \cite{preskill2018quantum}. Therefore, it is a highly challenging task to design applications with moderate quantum resources requirements which can leverage the quantum advantages on these NISQ devices.

%
%
With the above-mentioned two rapid growing fields, it is then natural to consider the combination of them. Especially the machine learning applications which can be implemented on NISQ devices. Indeed, the area \emph{quantum machine learning} (QML) draws a lot of attention recently and there are several promising breakthroughs. The most notable progress is the development of variational algorithms \cite{peruzzo2014variational, cerezo2020variational,bharti2021noisy} which enable the quantum machine learning on NISQ devices \cite{mitarai2018quantum}. Recent efforts have demonstrated the promising application of NISQ devices in several machine learning tasks \cite{ schuld2018circuit, Farhi2018ClassificationProcessors, benedetti2019parameterized, mari2019transfer, abohashima2020classification, easom2020towards, sarma2019quantum, chen2020hybrid, stein2020hybrid,chen2020quantum,kyriienko2020solving,dallaire2018quantum,li2021quantum, stein2020qugan, zoufal2019quantum, situ2018quantum,nakaji2020quantum,lloyd2020quantum, nghiem2020unified,chen19, lockwood2020reinforcement,wu2020quantum, jerbi2019quantum, Chih-ChiehCHEN2020,bausch2020recurrent,yang2020decentralizing}.
%

One of the common features of these successful ML models is that they are data-driven. To build a successful deep learning model, it requires a huge amount of data. Although there are several public datasets for research purpose, most advanced and personalized models largely depend on the collected data from users' mobile devices and other personal data (e.g. medical record, browsing habits and etc). For example, ML/DL approaches also succeed in the field of medical imaging \cite{suzuki2017overview,lundervold2019overview}, speech recognition \cite{deng2013new,amodei2016deep,hannun2014deep}, to name a few. These fields rely critically on the massive dataset collected from the population and these data should not be accessed by unauthorized third-party. The use of these sensitive and personally identifiable information raises several concerns. 
One of the concerns is that the channel used to exchange with the cloud service providers can be compromised, leading to the leakage of high-value personal or commercial data. Even if the communication channel can be secured, the cloud service provider is also risky as malicious adversaries can potentially invade the computing infrastructure. There are several solutions to deal with such issues. One of them is called \emph{federated learning} (FL), which focuses on the decentralized computing architecture. For example, users can train a speech recognition model on his cell phone and upload the model to the cloud in exchange of the global model without upload the recordings directly. Such framework is made possible due to the fact of recent advances in hardware development, making even the small devices so powerful.
This concept not only help the privacy-preserving practice in classical machine learning but also in the rapid emerging \emph{quantum machine learning} as researchers are trying to expand the machine learning capabilities by leveraging the power of quantum computers.
To harvest the power of quantum computers in the NISQ era, the key challenge is how to distribute the computational tasks to different quantum machines with limited quantum capabilities. Another challenge is the rising privacy concern in the use of large scale machine learning infrastructure. We address these two challenges by providing the framework of training quantum machine learning models in a federated manner.

In this paper, we propose the federated training on hybrid quantum-classical classifiers. We show that with the federated training, the performance in terms of the testing accuracy does not decrease. In addition, the model still converges quickly compared to the non-federated training. Our efforts not only help building secure QML infrastructure but also help the distributed QML training which is to better utilize available NISQ devices.

This paper is organized as follows. In Section~\ref{sec:FederatedMachineLearning}, we introduce the concept of federated machine learning. In Section~\ref{sec:VariationalQuantumCircuits}, we describe the variational quantum circuit architecture in details. In Section~\ref{sec:HybridQuantumClassicalTransferLearning}, we describe the transfer learning in hybrid quantum-classical models. Section~\ref{sec:ExpAndResults} shows the performance of the proposed federated quantum learning on the experimental data, followed by further discussions in Section~\ref{sec:Discussion}. Finally we conclude in Section~\ref{sec:Conclusion}.

\section{\label{sec:FederatedMachineLearning}Federated Machine Learning}
\begin{figure}[htbp]
\centering
\includegraphics[width=1.0\linewidth]{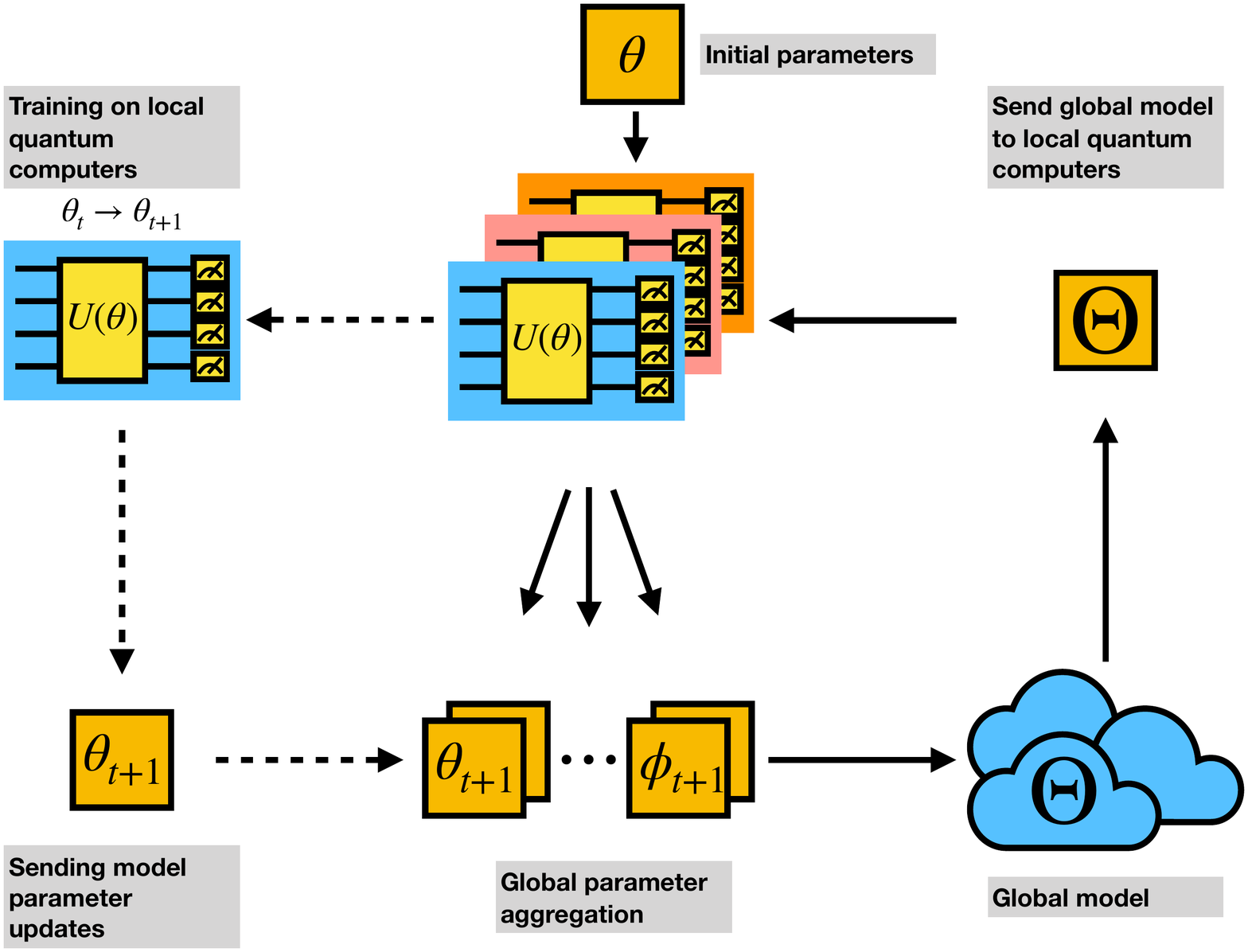}
\caption{{\bfseries Federated Quantum Machine Learning.}}
\label{fig:concept_federated_quantum_machine_learning}
\end{figure}

Federated learning (FL) \cite{mcmahan2017communication} emerges recently along with the rising privacy concerns in the use of large-scale dataset and cloud-based deep learning \cite{shokri2015privacy}. The basic components in a federated learning process are a \emph{central node} and several \emph{client nodes}. The central node holds the \emph{global model} and receives the trained parameters from client devices. The central node performs the \emph{aggregation} process to generate the new global model and share this new model to all of its client nodes. The client nodes will train locally with the received model with their own part of data, which in general is only a small portion.
In our proposed framework, the local clients are quantum computers or quantum simulators with the circuit parameters trained via hybrid quantum-classical manner. In each training round, a specified number of client nodes will be selected to perform the local training. Once the client training is finished, the circuit parameters from all the client nodes will be aggregated by the central node. There are various methods to aggregate the model. In this work, we choose the \emph{mean} of the client models.
The scheme of federated quantum machine learning is shown in \figureautorefname{\ref{fig:concept_federated_quantum_machine_learning}}.
For further discussion and advanced settings on federated learning, we refer to \cite{kulkarni2020survey, kairouz2019advances, lim2020federated, yang2019federated, li2020federated, li2019federated,wang2020convergence,semwal2020fedperf}.

\section{\label{sec:VariationalQuantumCircuits}Variational Quantum Circuits}
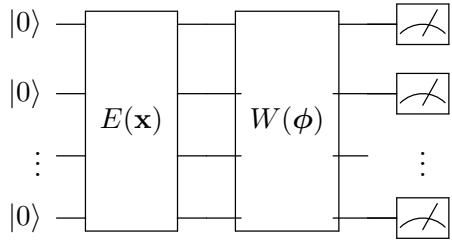
\begin{figure}[htbp]
\begin{center}
\begin{minipage}{10cm}
\Qcircuit @C=1em @R=1em {
\lstick{\ket{0}} & \multigate{3}{E(\mathbf{x})}  & \qw        & \multigate{3}{W(\boldsymbol{\phi})}       & \qw      & \meter \qw \\
\lstick{\ket{0}} & \ghost{F(\mathbf{x})}         & \qw        & \ghost{V(\boldsymbol{\theta})}              & \qw      & \meter \qw \\
\lstick{\vdots} & \ghost{F(\mathbf{x})}         & \qw        & \ghost{V(\boldsymbol{\theta})}              & \qw      &  \vdots \\
\lstick{\ket{0}} & \ghost{F(\mathbf{x})}         & \qw        & \ghost{V(\boldsymbol{\theta})}              & \qw      & \meter \qw \\
}
\end{minipage}
\end{center}
\caption[General structure for the variational quantum circuit.]{{\bfseries General structure for the variational quantum circuit (VQC).}
The $E(\mathbf{x})$ is the quantum routine for encoding the classical data into the quantum state and $W(\boldsymbol{\theta})$ is the variational quantum circuit block with the learnable parameters $\boldsymbol{\phi}$. After the quantum operation, the quantum state is \emph{measured} to retrieve classical numbers for further processing.
}
\label{Fig:GeneralVQC}
\end{figure}

Variational quantum circuits (VQC) or quantum neural networks (QNN) are a special kind of quantum circuits with adjustable circuit parameters subject to optimization procedures developed by the classical machine learning community. In \figureautorefname{\ref{Fig:GeneralVQC}} we introduce the general setting of a VQC in which the $E(\mathbf{x})$ encodes the classical data into a quantum state which the quantum gates can actually operate on and the $W(\phi)$ is the learnable block which can be seen as the \emph{weights} in classical neural network. There are \emph{quantum measurements} in the final part of VQC which is to \emph{readout} the information from a quantum circuit and these classical numbers can be further processed with other classical or quantum components.
The general idea of VQC or QNN is that the circuit parameters are updated via iterative methods on a classical computer. 
Recent theoretical studies have also demonstrated that VQCs are more expressive than conventional neural networks~\cite{sim2019expressibility,lanting2014entanglement,du2018expressive, abbas2020power} with respect to the number of parameters or the learning speed. In addition, in the work \cite{chen2020quantum} and \cite{chen2020qcnn, chen2021hybrid}, it has been demonstrated via numerical simulation that certain hybrid quantum-classical architectures reach higher accuracies than classical neural networks with similar number of parameters.
\begin{figure}[htbp]
\begin{center}
\begin{minipage}{10cm}
\Qcircuit @C=1em @R=1em {
\lstick{\ket{0}} & \gate{R_y(\arctan(x_1))} & \gate{R_z(\arctan(x_1^2))} & \ctrl{1}   & \qw       & \qw      & \targ   & \gate{R(\alpha_1, \beta_1, \gamma_1)} & \meter \qw \\
\lstick{\ket{0}} & \gate{R_y(\arctan(x_2))} & \gate{R_z(\arctan(x_2^2))} & \targ      & \ctrl{1}  & \qw      & \qw     & \gate{R(\alpha_2, \beta_2, \gamma_2)} &  \meter \qw \\
\lstick{\ket{0}} & \gate{R_y(\arctan(x_3))} & \gate{R_z(\arctan(x_3^2))} & \qw        & \targ     & \ctrl{1} & \qw     & \gate{R(\alpha_3, \beta_3, \gamma_3)} &  \qw \\
\lstick{\ket{0}} & \gate{R_y(\arctan(x_4))} & \gate{R_z(\arctan(x_4^2))} & \qw        & \qw       & \targ    & \ctrl{-3}& \gate{R(\alpha_4, \beta_4, \gamma_4)} & \qw \gategroup{1}{4}{4}{8}{.7em}{--}\qw 
}
\end{minipage}
\end{center}
\caption[Variational quantum classifier.]{{\bfseries Variational quantum classifier.}
  The variational quantum classifier includes three components: \emph{encoder}, \emph{variational layer} and \emph{quantum measurement}. The encoder consists of several single-qubit gates $R_y(\arctan(x_i))$ and $R_z(\arctan(x_i^2))$ which represent rotations along $y$-axis and $z$-axis by the given angle $\arctan(x_i)$ and $\arctan(x_i^2)$, respectively. These rotation angles are derived from the input values $x_i$ and are not subject to iterative optimization.
  The variational layer consists of CNOT gates between each pair of neighbouring qubits which are used to entangle quantum states from each qubit and general single qubit unitary gates $R(\alpha,\beta,\gamma)$ with three parameters $\alpha,\beta,\gamma$. Parameters labeled $\alpha_i$, $\beta_i$ and $\gamma_i$ are the ones for iterative optimization. The quantum measurement component will output the Pauli-$Z$ expectation values of designated qubits. 
  The number of qubits and the number of measurements can be adjusted to fit the problem of interest. In this work, we use the VQC as the final classifier layer, therefore the number of qubits equals to the latent vector size which is $4$ and we only consider the measurement on the first two qubits for binary classification. The grouped box in the VQC may repeat several times to increase the number of parameters, subject to the capacity and capability of the available quantum computers or simulation software used for the experiments. In this work, the grouped box repeats for $2$ times.}

\label{Fig:Basic_VQC}
\end{figure}
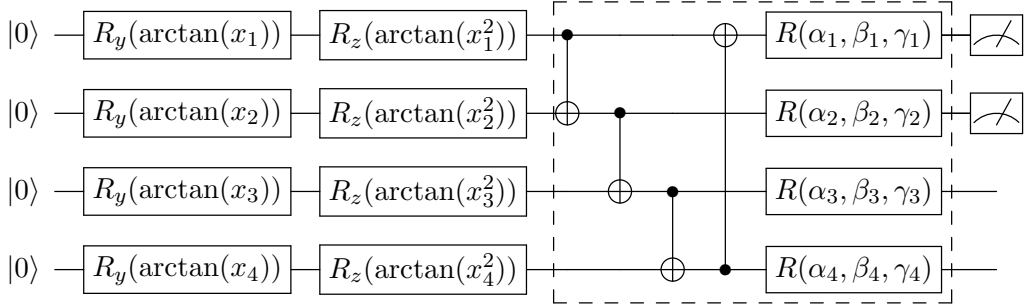

Recent advances in VQC have demonstrated various applications in a wide variety of machine learning tasks. For example, VQC has been shown to be successful in the task of classification \cite{mitarai2018quantum, schuld2018circuit, Farhi2018ClassificationProcessors, benedetti2019parameterized, mari2019transfer, abohashima2020classification, easom2020towards, sarma2019quantum, chen2020hybrid, stein2020hybrid,sierra2020dementia, chen2020qcnn,wu2020application}, function approximation \cite{chen2020quantum, mitarai2018quantum,kyriienko2020solving}, generative machine learning \cite{dallaire2018quantum, stein2020qugan, zoufal2019quantum, situ2018quantum,nakaji2020quantum,li2021quantum}, metric learning \cite{lloyd2020quantum, nghiem2020unified}, deep reinforcement learning \cite{chen19, lockwood2020reinforcement, jerbi2019quantum, Chih-ChiehCHEN2020,wu2020quantum,jerbi2021variational}, sequential learning \cite{chen2020quantum, bausch2020recurrent, takaki2020learning} and speech recognition \cite{yang2020decentralizing}.
\subsection{Quantum Encoder}
For a quantum circuit to operate on a classical dataset, the critical step is to define the \emph{encoding} method which is to transform the classical vector into a quantum state. The encoding scheme is important as it is relevant to the efficiency of hardware implementation and potential quantum advantages. In NISQ era, the number of qubits as well as the circuit-depth are limited. Therefore, we need to encode the classical values with small number of qubits and without too many quantum operations. For more in-depth introduction of various kinds of encoding methods used in QML, refer to \cite{Schuld2018InformationEncoding}.
A general $N$-qubit quantum state can be represented as:
\begin{equation}
\label{eqn:quantum_state_vec}
    \ket{\psi} = \sum_{(q_1,q_2,...,q_N) \in \{ 0,1\}^N}^{} c_{q_1,...,q_N}\ket{q_1} \otimes \ket{q_2} \otimes \ket{q_3} \otimes ... \otimes \ket{q_N},
\end{equation}
where $ c_{q_1,...,q_N} \in \mathbb{C}$ is the \emph{amplitude} of each quantum state and $q_i \in \{0,1\}$. 
The square of the amplitude $c_{q_1,...,q_N}$ is the \emph{probability} of measurement with the post-measurement state in  $\ket{q_1} \otimes \ket{q_2} \otimes \ket{q_3} \otimes ... \otimes \ket{q_N}$, and the total probability should sum to $1$, i.e.,
\begin{equation} 
\label{eqn:quantum_state_vec_normalization_condition}
\sum_{(q_1,q_2,...,q_N) \in \{ 0,1\}^N}^{} ||c_{q_1,...,q_N}||^2 = 1. 
\end{equation}
In this work, we use use the \emph{variational encoding} scheme to encode the classical values into a quantum state. The basic idea behind this encoding scheme is to use the input values or their transformation as rotation angles for the quantum rotation gate. As shown in \figureautorefname{\ref{Fig:Basic_VQC}}, the encoding parts consist of single-qubit rotation gates $R_y$ and $R_z$ and use $\arctan(x_i)$ and $\arctan(x_i^{2})$ as the corresponding transformations.
\subsection{Quantum Gradients}
The hybrid quantum-classical model can be trained in an end-to-end fashion, following the common backpropagation method used in training deep neural network. When it comes to the gradient calculation on quantum functions, \emph{parameter-shift} method is employed. It can be used to derive the analytical gradient of the quantum circuits. The method is described in the reference~\cite{schuld2019evaluating,bergholm2018pennylane}. The idea behind the parameter-shift rule is that given the knowledge of calculating the expectation of certain observable of quantum functions, the quantum gradients can be calculated without the use of finite difference method.

\section{\label{sec:HybridQuantumClassicalTransferLearning}Hybrid Quantum-Classical Transfer Learning}
\begin{figure}[htbp]
\centering
\includegraphics[width=1.0\linewidth]{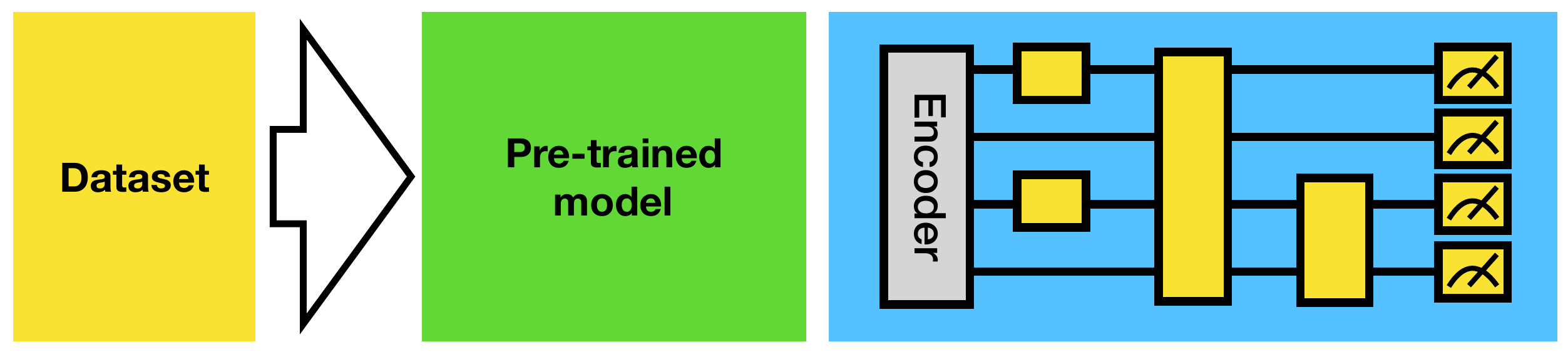}
\caption{{\bfseries Hybrid Quantum-Classical Transfer Learning.}}
\label{fig:concept_hybrid_transfer_learning}
\end{figure}

%
In the NISQ era, the quantum computers are not error-corrected and thus cannot perform calculations in a fault-tolerant manner. The circuit depth and number of qubits are therefore limited and it is non-trivial to design the model architectures which can potentially harness the capabilities provided by near-term quantum computers.
In this work, we employ the hybrid quantum-classical transfer learning scheme inspired by the work \cite{mari2019transfer}. The idea is to use a pre-trained classical deep neural network, mostly convolutional neural networks (CNN) to extract the features from the images and compress the information into a latent vector $x$ which is with much smaller dimension than the original image. Then the latent vector $x$ is processed by the quantum circuit model to output the logits of each class. The scheme is presented in \figureautorefname{\ref{fig:concept_hybrid_transfer_learning}}. In this work, we employ the VGG16 \cite{simonyan2014very} pre-trained model as the feature extractor.
\section{\label{sec:ExpAndResults}Experiments and Results}
In this study we consider the following setting:
\begin{itemize}
    \item Central node $C$: Receive the uploaded circuit parameters $\theta_i$ from each local machine $N_i$ and aggregate them into a global parameter $\Theta$ and distributes to all local machines.
    \item Training points are equally distributed to the local machines and the testing points are on the central node to evaluate the aggregated global model.
    \item Individual local machines $N_i$: Each has a distinct part of the training data and will perform $E$ epochs of the training locally with the batch size $B$.
    
\end{itemize}

The software we use for this work are PyTorch \cite{paszke2019pytorch}, PennyLane \cite{bergholm2018pennylane} and Qulacs \cite{suzuki2020qulacs}.
\subsection{Cats vs Dogs}
\begin{figure}[htbp]
\centering
\includegraphics[width=1.0\linewidth]{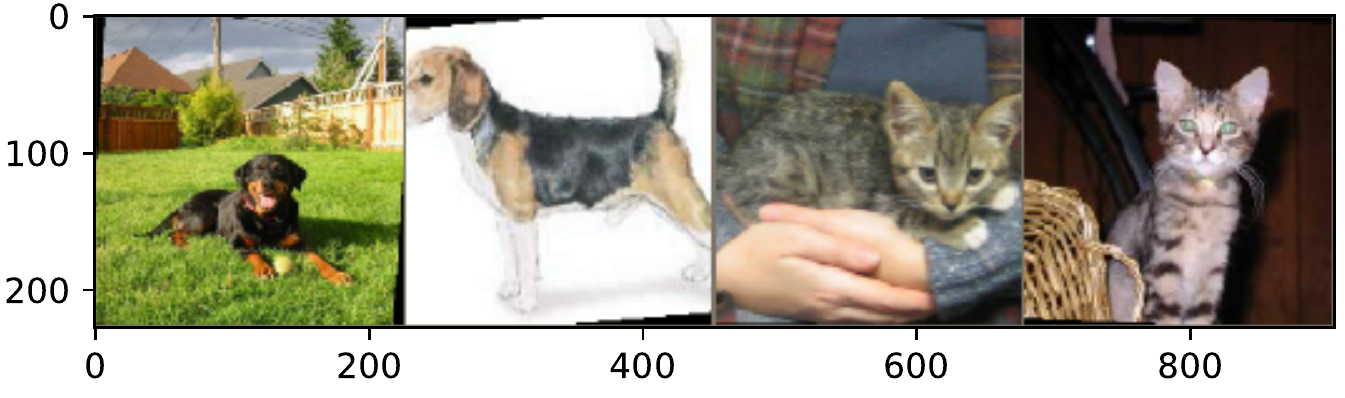}
\caption{{\bfseries Cats vs Dogs Dataset.} \cite{asirra-a-captcha-that-exploits-interest-aligned-manual-image-categorization}}
\label{fig:demo_cats_dogs}
\end{figure}

\begin{figure}[htbp]
\centering
\includegraphics[width=1.0\linewidth]{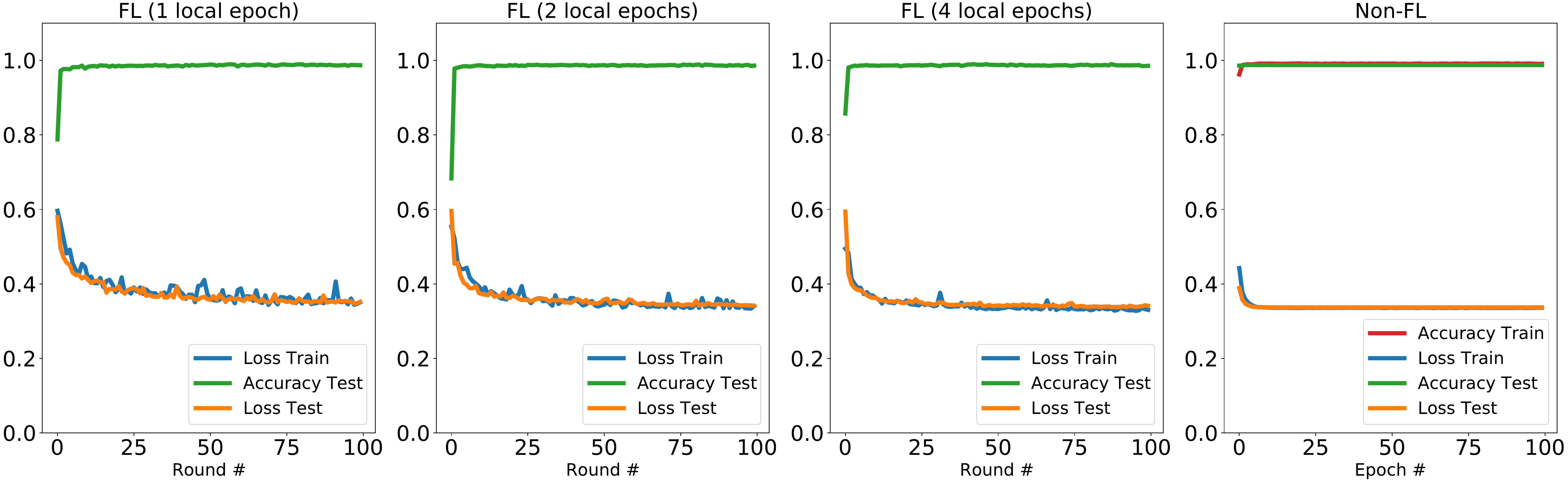}
\caption{{\bfseries Results: Cats vs Dogs.}}
\label{fig:result_cats_dogs}
\end{figure}

We perform the binary classification on the classic cats vs dogs dataset \cite{asirra-a-captcha-that-exploits-interest-aligned-manual-image-categorization}. Each image in this dataset has slightly different dimensions, therefore we preprocessed to make all of the training and testing samples in the dimension of $224 \times 224$. In \figureautorefname{\ref{fig:demo_cats_dogs}} we show some of the examples from this dataset.
%
Here, we have in total 23,000 training data and 2,000 testing data. The testing data are on the central node which will be used to evaluate the aggregated model (global model) after each training round. The training data are equally distributed to the $100$ local machines $N_i$ where $i \in \{1 \cdots 100\}$. Therefore in each local machine, there are $230$ training points. In each training round, $5$ local machines will be randomly selected and each will perform $1$, $2$ or $4$ epochs of training with its own training points. The batch size is $S = 32$. The trained model will then be sent to the central node for the aggregation. The aggregation method we use in this experiment is the collected model average. The aggregated model (global model) will then be shared to all the local machines. 
The hybrid model used in this experiment consists of pre-trained VGG16 model and a $4$-qubit variational quantum circuit (VQC) as shown in \figureautorefname{\ref{Fig:Basic_VQC}}. The original classifier layer in the VGG16 model is replaced with the one shown in \tableautorefname{\ref{tab:classifier_vgg}} in order to fit input dimension of the VQC layer. The dashed-box in the quantum circuit repeats twice, consisting of $4 \times 3 \times 2 = 24$ quantum circuit parameters. The VQC receives $4$-dimensional compressed vectors from the pre-trained VGG model to perform the classification task. 
The non-federated training for the comparison with the same hybrid VGG-VQC architecture as the one used in the federated training.
We perform $100$ training rounds and the results are presented in \figureautorefname{\ref{fig:result_cats_dogs}}. We compare the performance of federated learning with non-federated learning with the same hybrid quantum-classical architecture and the same dataset. 
\begin{table}[htbp]
\begin{tabular}{|l|l|l|l|l|l|l|l|}
\hline
       & Linear & ReLU        & Dropout($p = 0.5$)       & Linear & ReLU        & Dropout($p = 0.5$)       & Linear \\ \hline
Input  & 25088  & \multicolumn{2}{l|}{\multirow{2}{*}{}} & 4096   & \multicolumn{2}{l|}{\multirow{2}{*}{}} & 4096   \\ \cline{1-2} \cline{5-5} \cline{8-8} 
Output & 4096   & \multicolumn{2}{l|}{}                  & 4096   & \multicolumn{2}{l|}{}                  & 4      \\ \hline
\end{tabular}
\caption{The trainable layer in our modified VGG model.}
\label{tab:classifier_vgg}
\end{table}

%
In the left three panels of \figureautorefname{\ref{fig:result_cats_dogs}}, we present the results of training the hybrid quantum model via federated setting with different number of local training epochs. Since the training data is distributed across different clients, we only consider the testing accuracies with the aggregated global model. In the considered Cats vs Dogs dataset, we observe that both the testing accuracies and testing loss reach the comparable level as the non-federated training. We also observe that the training loss, which is the average from clients, has fluctuations compared to non-federated training (shown in \tableautorefname{\ref{tab:comparison_performance_cats_dogs}}). The underlying reason might be that in each training round, different clients are selected, therefore the training data used to evaluate the training loss are different. Yet the training loss still converges after the $100$ rounds of training.
In addition, the testing loss and accuracies converge to comparable levels to the non-federal training, regardless of the local training epochs. Notably, we observe that a single epoch in local training is pretty enough to train a well-performed model. In each round of the federated training, the model updates are based on the samplings from $5$ clients, with $1$ local training epoch. The computing resources used are linear with $230 \times 5 \times 1 = 1150$ in total. While for a full epoch of training with non-federated setting, the computing resources used are linear with $23000$. This results imply the potential of more efficient training on QML models with distributed schemes. This particularly benefits the training of quantum models when we are using high-performance simulation platform or an array of small NISQ devices, with the communication overhead is moderate.
\begin{table}[htbp]
\centering
\begin{tabular}{|l|l|l|l|}
\hline
                       & Training Loss & Testing Loss & Testing Accuracy \\ \hline
Federated Training (1 local epoch)     & $0.3506$      & $0.3519$     & $98.7\%$   \\ \hline
Federated Training (2 local epochs)     & $0.3405$      & $0.3408$     & $98.6\%$   \\ \hline
Federated Training (4 local epochs)     & $0.3304$      & $0.3413$     & $98.6\%$   \\ \hline
Non-Federated Training & $0.3360$      & $0.3369$     & $98.75\%$              \\ \hline
\end{tabular}
\caption{Comparison of performance in different training schemes with Cats vs Dogs dataset.}
\label{tab:comparison_performance_cats_dogs}
\end{table}
\subsection{CIFAR (Planes vs Cars)}
\begin{figure}[htbp]
\centering
\includegraphics[width=1.0\linewidth]{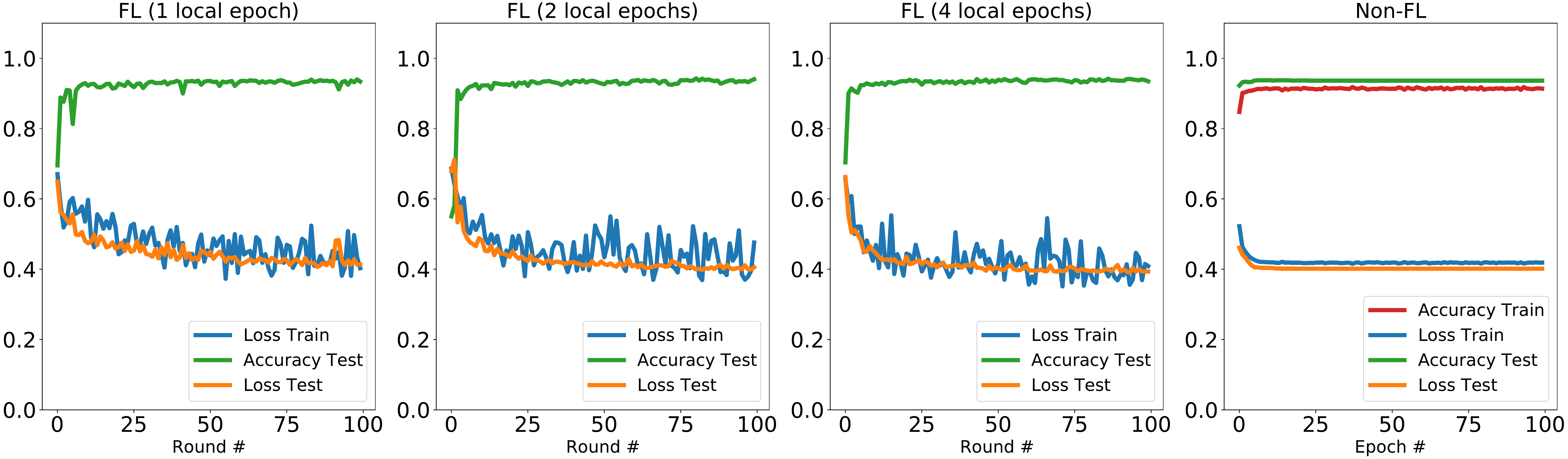}
\caption{{\bfseries Results: Planes vs Cars.} }
\label{fig:result_planes_cars}
\end{figure}

In this experiment, we use the data from the CIFAR-10 dataset \cite{Krizhevsky09learningmultiple}. The dimension of the images in this dataset is $32 \times 32$.
The hybrid quantum-classical models used in this experiment is the same as in the previous experiment.
Here we have in total $10000$ training points and $2000$ testing points. The testing points are on the central node (global model) which will be used to evaluate the aggregated model after each training round. The training points are equally distributed to the $100$ local machines $N_i$ where $i \in \{1 \cdots 100\}$. In each training round, $5$ local machines will be randomly selected and each will perform $1$, $2$ or $4$ epochs of training with its own training points. The batch size $S = 32$. The trained model will then be sent to the central mode for aggregation.
%
%
In the left three panels of \figureautorefname{\ref{fig:result_planes_cars}}, we present the results of training the hybrid quantum model via federated setting with different number of local training epochs. Since the training data is distributed across different clients, we only consider the testing accuracies with the aggregated global model. In the considered Planes vs Cars dataset, we observe that both the testing accuracies and testing loss reach the comparable level as the non-federated training (shown in \tableautorefname{\ref{tab:comparison_performance_planes_cars}}). Similar to the previous Cats vs Dogs dataset, we observe that the training loss, which is the average from clients, has fluctuations compared to non-federated training. 
%
In addition, the testing loss and accuracies converge to comparable levels to the non-federated training, regardless of the local training epochs.
Notably, we observe that a single epoch in local training is pretty enough to train a well-performed model. In each round of the federated training, the model updates are based on the samplings from $5$ clients, with $1$ local training epoch. The computing resources used are linear with $100 \times 5 \times 1 = 500$ in total. While for a full epoch of training with non-federated setting, the computing resources used are linear with $10000$. This results again imply the potential of more efficient training on QML models with distributed schemes. This particularly benefits the training of quantum models when we are using high-performance simulation platform or an array of small NISQ devices, with the communication overhead is moderate.
\begin{figure}[htbp]
\centering
\includegraphics[width=1.0\linewidth]{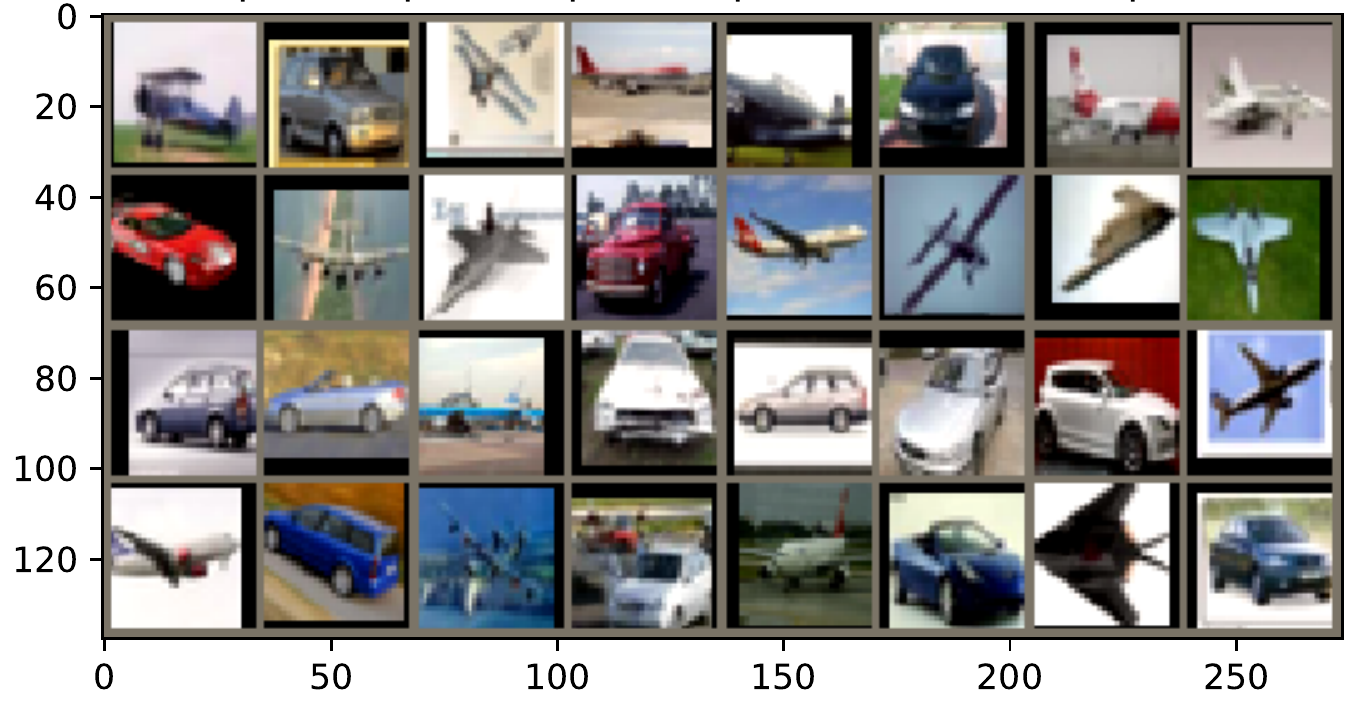}
\caption{{\bfseries Planes vs Cars from CIFAR-10 Dataset.} \cite{Krizhevsky09learningmultiple}}
\label{fig:demo_planes_cars}
\end{figure}
\begin{table}[htbp]
\centering
\begin{tabular}{|l|l|l|l|}
\hline
                       & Training Loss & Testing Loss & Testing Accuracy \\ \hline
Federated Training (1 local epoch)     & $0.4029$      & $0.4133$     & $93.4\%$   \\ \hline
Federated Training (2 local epochs)     & $0.4760$      & $0.4056$     & $94.05\%$  \\ \hline
Federated Training (4 local epochs)     & $0.4090$      & $0.3934$     & $93.45\%$   \\ \hline
Non-Federated Training & $0.4190$      & $0.4016$     & $93.65\%$               \\ \hline
\end{tabular}
\caption{Comparison of performance in different training schemes with CIFAR (Planes vs Cars) dataset.}
\label{tab:comparison_performance_planes_cars}
\end{table}
%



\section{\label{sec:Discussion}Discussion}
\subsection{Integration with Other Private-Preserving Protocols}
In this study we consider the federated quantum learning framework. One of the limitation is that the process of exchanging model parameters can potentially be attacked. Moreover, we can not exclude the possibilities that malicious parties are joining the network, which will get the aggregated global model. The leaked model parameters can be used to deduce the training data of the model \cite{dwork2014algorithmic}. 
There are other protocols which can further boost the security. For example, it has been shown that trained models can be used to recover training entries \cite{fredrikson2015model}. In addition, it is also possible for adversaries to find out whether a specific entry is used in training process \cite{shokri2017membership}. These possibilities raise serious concerns when the QML models are used to process private and sensitive data.  One of the potential solution is to train the model with \emph{differential privacy} (DP) \cite{abadi2016deep}. With DP, it is possible to share the trained model and still keep the private information of the training data. Another direction if to incorporate the secure multi-party computation \cite{goryczka2013secure} which can further increase the security in decentralization computing.
\subsection{Different Aggregation Method}
In this work, we use the simplest aggregation method which is simply calculating the average of parameters from each local machine (model). In a more realistic application scenario, clients may upload a corrupt trained model, or the communication channel may be interfered by noise, which can potentially compromise the global model if there is no other countermeasures. Several recent works present advanced aggregation schemes to address this issue \cite{pillutla2019robust, ang2020robust}. The implementation of these advanced protocols with quantum machine learning is an interesting direction for future work.
\subsection{Decentralization}
This research presents a proof-of-concept federated training on quantum machine learning models. The scheme includes a \emph{central node} to receive the trained models from clients, to aggregate them and to distribute the aggregated model to clients. This \emph{central node} can be vulnerable to malicious attacks and the adversaries can compromise the whole network. Moreover, the communication bandwidth between clients and the central node may vary, leading some undesired effects in the synchronization process. To address these issues, recent studies propose various \emph{decentralized} federated learning schemes \cite{savazzi2020federated, wittkopp2021decentralized, pokhrel2020decentralized, bonawitz2019towards, xiao2020fully, lalitha2018fully, lu2020decentralized}.
%
For example, the \emph{distributed ledger technologies} (DLT) \cite{nakamoto2008bitcoin, zyskind2015decentralizing, cai2018decentralized, pandl2020convergence} which power the development of blockchain have been applied in the decentralized FL \cite{qu2020decentralized, ramanan2020baffle, awan2019poster, qu2020blockchained, zhao2020privacy, bao2019flchain, kim2019blockchained}. The blockchain technologies are used to ensure the robustness and integrity of the shared information while remove the requirement of a central node. Blockchain-enabled FL can also be designed to encourage the data-owner participating in the model training process \cite{liu2020fedcoin}. 
In addition, \emph{peer-to-peer} protocols are also employed in FL to remove the need of a central node \cite{roy2019braintorrent, lalitha2019peer}. \emph{Gossip learning} \cite{hegedHus2016robust, ormandi2013gossip} is an alternative learning framework to FL \cite{hegedHus2019gossip, hegedHus2019decentralized, hegedHus2021decentralized, hu2019decentralized}. Under gossip learning framework, no central node is required, nodes on the network exchange and aggregate models directly.
The efficiencies and capabilities of these decentralized schemes such as blockchained FL and gossip learning in the quantum regime are left for future work. 

In classical machine learning, distributed training frameworks are designed to scale up the model training to computing clusters \cite{sergeev2018horovod}, making the training on large-scale dataset and complex models possible.
Potential direction is to apply the federated quantum learning to the high-performance quantum simulation.  
\subsection{Other Quantum Machine Learning Models}
In this work we consider the hybrid quantum-classical transfer learning architecture which includes a pre-trained classical model as the feature extractor. Currently the available quantum computers and simulation software are rather limited and do not posses large number of qubits. However, the proposed framework can be extended well beyond the transfer learning structure. Recently, a hybrid architecture combining tensor network and quantum circuit is proposed \cite{chen2021end}. Such hybrid architecture is more generic than the pre-trained network used in this work. It is interesting to investigate the potential of decentralizing such kind of architectures. Moreover, it is possible to study the federated learning on quantum convolutional neural networks (QCNN) \cite{chen2020qcnn,cong2019quantum, li2020quantum, oh2020tutorial, kerenidis2019quantum, liu2019hybrid,chen2021hybrid} when larger-scale quantum simulators or real quantum computers are available.
\subsection{Potential Applications}
This work can potentially be integrated with the work \cite{yang2020decentralizing,qi2020submodular} for decentralizing the quantum-enhanced speech recognition. Another potential direction is in the use in healthcare in which a tremendous amount of sensitive personal data need to be processed to train a reliable model. For example, the work \cite{sierra2020dementia} studied the application of VQC in dementia prediction which would benefit from the federated training to preserve the users' privacy. Recently, the application of quantum computing in financial industries have drawn a lot of attention \cite{egger2020quantum}. It is expected that federated QML would play an important role in finance as well.
\section{\label{sec:Conclusion}Conclusion}
In this work we provide the framework to train hybrid quantum-classical classifiers in a federated manner which can help in preserving the privacy and distributing computational loads to an array of NISQ computers. We also show that the federated training in our setting does not sacrifice the performance in terms of the testing accuracy. This work should benefit the research in both the privacy-preserving AI and the quantum computing and pave new direction on building secure, reliable and scalable distributed quantum machine learning architecture.

\begin{acknowledgments}
This work is supported by the U.S.\ Department of Energy, Office of Science, Advanced Scientific Computing Research under Award Number DE-SC-0012704 and the Brookhaven National Laboratory LDRD \#20-024.
\end{acknowledgments}

\appendix


\bibliographystyle{ieeetr}
\bibliography{apssamp,bib/tool,bib/qml_examples,bib/classical_ml,bib/qcnn,bib/fed_ml,bib/vqc,bib/qc,bib/application,bib/dp,bib/decentralized_fl}

\end{document}